\documentclass[prl,twocolumn,showpacs]{revtex4}
\usepackage{graphicx}
\usepackage{amsfonts}
\begin{document}

\title{Predict-prevent control method for perturbed excitable systems}

\author{Marzena Ciszak$^1$, Claudio R. Mirasso$^{2}$, Ra\'ul Toral$^{2}$, Oscar Calvo$^{3}$}
\affiliation{ $^1$C.N.R.-Istituto Nazionale di Ottica Applicata,
L.go E. Fermi 6, 50125 Florence, Italy \\
$^2$IFISC (Instituto de
F\'{\i}sica Interdisciplinar y Sistemas Complejos), CSIC-UIB, Campus
UIB, E-07122 Palma de Mallorca, Spain\\
$^3$Departament de F{\'\i}sica, Universitat de les Illes Balears, E-07122 Palma de Mallorca, Spain
 }

\begin{abstract}
We present a control method based on two steps: prediction and prevention. For prediction we use the anticipated synchronization scheme, considering unidirectional coupling between excitable systems in a master-slave configuration. The master is the perturbed system to be controlled, meanwhile the slave is an auxiliary system which is used to predict the master's behavior. We demonstrate theoretically and experimentally that an efficient control may be achieved.
\end{abstract}

\date{\today}
\maketitle

A classical problem in engineering science is the control of dynamical systems~\cite{control}. While earliest applications in this field aimed to inhibit the instabilities in electronic devices, control studies turned, later on, into other areas. Of particular interest are the applications to medicine since transitions from regular to irregular oscillations of many organs of the human body have been found to be associated with a diseased behavior. As an example, the heart beat has been found to undergo chaotic behavior during arrhythmias and other heart diseases~\cite{heart}.

One of the best known methods for stabilizing unstable periodic orbits is the one introduced by Ott, Grebogi and Yorke (OGY)~\cite{ogy}. It uses a small perturbation in some parameter of the system to stabilize an unstable orbit into a periodic one. The delayed-feedback control (DFC), introduced by Pyragas~\cite{pyragas}, intends the same but uses a self-feedback loop and works when the feedback delay is close to the period of the unstable orbit that one aims to stabilize. It has been experimentally implemented in optical systems~\cite{sukow}, chaotic flows~\cite{luth}, and cardiac systems~\cite{hall}, among others.

In this paper we present a rather different control method with particular application, but not necessarily limited, to excitable systems. These systems have been used to model the behavior of many cell types, including the typical spiking dynamics of heart cells and neurons. Excitable systems are characterized by a highly non-linear response to an external perturbation: if the perturbation is sufficiently small, the system returns smoothly to its steady state; if the perturbation exceeds a certain threshold, the system fires an excitable spike, sometimes called the action potential. During a refractory period following the spike, perturbations of moderate amplitude do not alter much the dynamics of the system. Although excitable systems can involve a large number of variables (as in the case of neurons), their essential features can be captured with a much-reduced description. We will be using in this paper the one-dimensional Adler and the two-variable FitzHugh-Nagumo systems as examples. They both provide some of the simplest representation of excitable dynamics.

The proposed control method relies on the phenomenon of {\sl anticipated synchronization}. Synchronization refers to the collective timing of coupled systems, from simple oscillators and clocks to some vital functions found in living organisms, among many others~\cite{PRK01,blekhman}. Anticipated synchronization refers to a particular regime, first studied by Voss~\cite{voss1}, which appears in unidirectionally coupled systems in a master-slave configuration. In this regime, two dynamical systems synchronize in such a way that the {\sl slave} system, ${\bf y}(t)$, anticipates the trajectory of the {\sl master}, ${\bf x}(t)$. One of the simplest schemes is based upon the general equations:
\begin{eqnarray}
\label{scheme1a}
\dot {\bf x}(t)& =& {\bf f}[{\bf x}(t)],\\
\label{scheme1b}
\dot {\bf y}(t) & = & {\bf f}[{\bf y}(t)]+{\bf K}[{\bf x}(t)-{\bf y}(t-\tau)].
\end{eqnarray}
The function ${\bf f}[{\bf x}]$ defines the dynamical system under consideration; ${\bf K}$ is a general coupling function satisfying ${\bf K}[{\bf 0}]={\bf 0}$ and $\tau$ is the delay time in the feedback loop of the slave. As proven by Voss~\cite{voss1}, the manifold \mbox{${\bf y}(t)={\bf x}(t+\tau)$}, i.e. the slave anticipates by a time $\tau$ the dynamics of the master, can be a (structurally) stable solution of equations (\ref{scheme1a}-\ref{scheme1b}). Anticipated synchronization has been studied theoretically and experimentally in many systems
~\cite{chialvo,M01,HMM02,our,our2,KHTM05,dynMar,LTDASL02,voss2,PJG08} (see~\cite{mplb} for a short review).

\begin{figure}
\begin{center}
\includegraphics[scale=0.8]{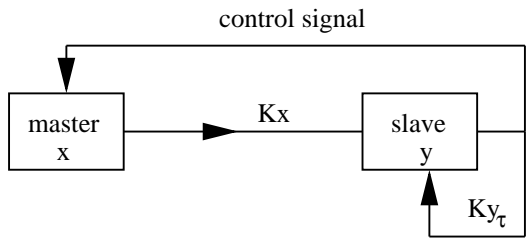}
\caption{\label{fig:scheme}Schematic presentation of the predict-prevent control method. The input signal $I(t)$ is applied only to the master system $\bf{x}$ while its influence on the slave $\bf{y}$ is indirectly through the unidirectional coupling $\bf{Kx}$. The slave dynamics has a delayed loop $\bf{Ky_{\tau}\equiv Ky(t-\tau)}$. The control signal is applied whenever a condition for the slave variable is satisfied.}
\end{center}
\end{figure}

Taking advantage of anticipated synchronization we propose a control method, which we apply to externally perturbed excitable systems, designed to correct unwanted dynamical behaviors. The method displays some of the most advantageous characteristics of OGY and DFC, but nevertheless differs essentially from them. The method uses the delayed feedback, as in the DFC method, but in an auxiliary slave system to process the information and decide whether to activate the control. After the decision of activating the control has been taken, a small perturbation is applied to the master in order to prevent its firing. This procedure resembles the perturbative action of OGY but it does not require any previous calculations. In the following we explain the new method in some detail. We show how it can be implemented to suppress unwanted spikes produced by the effect of external random perturbations. We prove that the control method can work under non-ideal conditions, such as the presence of noise in the dynamics and parameter mismatch between master and slave. Finally, we demonstrate experimentally the efficiency of the proposed method.

\begin{figure}
\centering
\includegraphics[scale=0.25,angle=-90]{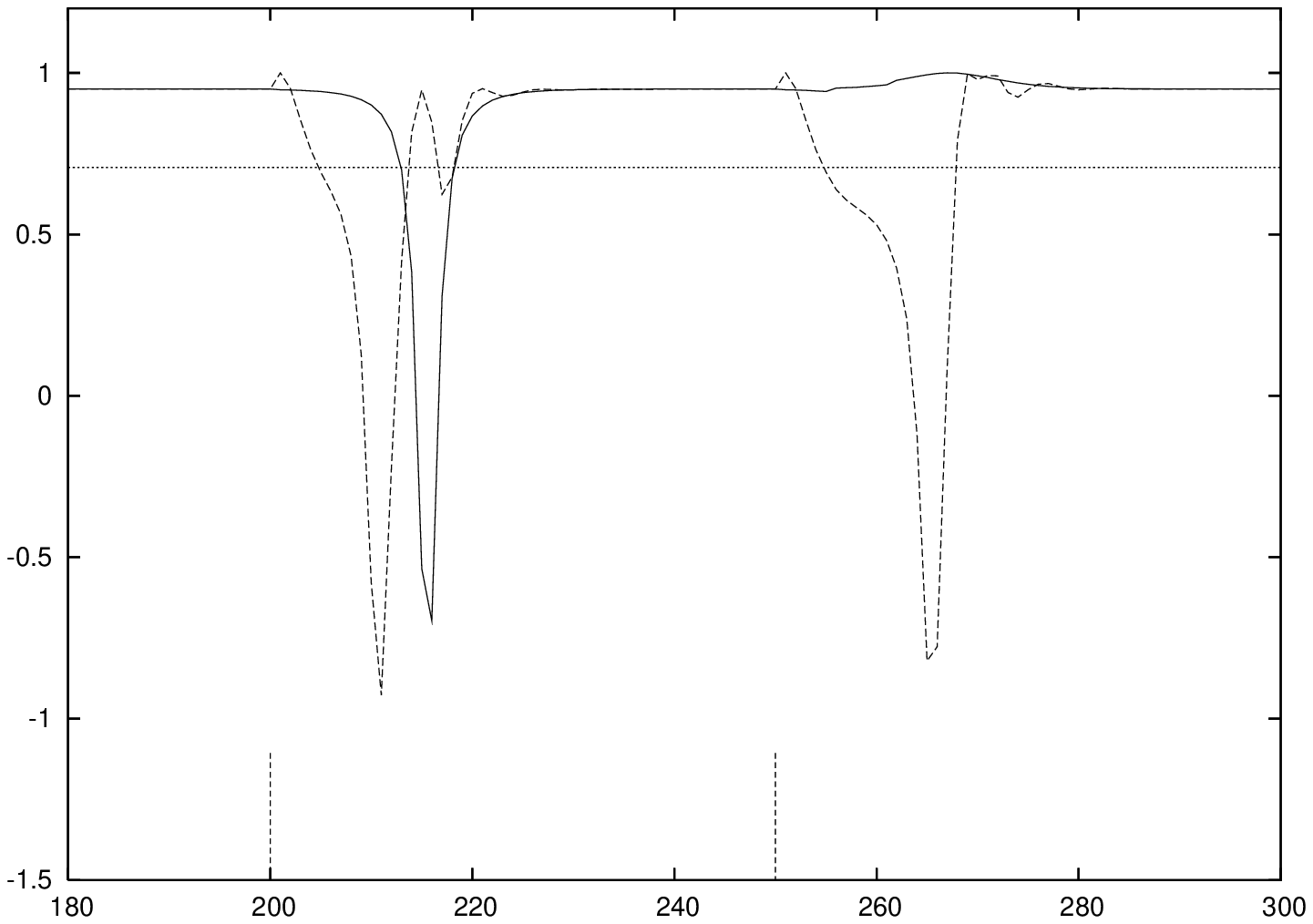}
\caption{ \label{fig:single}The solid curve plots $\cos[x(t)]$, the evolution of the master system, while the dotted line is $\cos[y(t)]$, the slave system coming from a numerical integration of Eqs.(\ref{ms1},\ref{ms2}). External perturbations of intensity $I_0=0.64$ are applied at times $t=200$ and $t=250$ (small vertical lines at the bottom). In the first perturbation ($t=200$) we see clearly that the slave anticipates the dynamics of the master. In the second perturbation ($t=250$) the spike of the master has been supressed by applying a correcting pulse of amplitude $\epsilon=0.04$ when the trajectory of the slave crosses the reference level $\cos[x_0]=0.707$ (horizontal line). We have used the parameters $\mu=0.95$, $K=0.6$, $\tau=1.5$ and a reaction time $t_R=1$.}
\end{figure}

Although our method is quite general, we consider, for the sake of concreteness, two unidirectionally coupled Adler systems~\cite{adler} in the master-slave configuration:
\begin{eqnarray}
\dot{x}(t)&=&\mu -\cos[x(t)]+I(t),\label{ms1}\\
\dot{y}(t)&=&\mu -\cos[y(t)]+K\sin[x(t)-y(t-\tau)]\label{ms2}.
\end{eqnarray}
$I(t)$ represents an external perturbation and $\mu$, $K$ and $\tau$ are constant parameters. Throughout the paper, we will take
the values $\mu=0.95$, $K=0.6$, $\tau=1.5$, which satisfy the conditions for anticipated synchronization~\cite{mplb,our2}. Note that the perturbation is applied only to the master system $x(t)$ while it acts upon the slave system $y(t)$ only indirectly through the coupling term. We use a $2\pi$ periodic function for coupling, appropriate for the angular-type variables considered. The slave is driven by the master and subject to its own feedback loop, of strength $K$, so producing the anticipated synchronization regime. The only modification with respect to Eqs.~(\ref{scheme1a}-\ref{scheme1b}) is the inclusion of the external perturbation $I(t)$ acting upon the dynamics of the master. As discussed in~\cite{dynMar}, the feedback term lowers the excitability threshold of the slave and allows it to react to the perturbation faster than the master. Hence, the slave is used to anticipate the future response of the master to this perturbation. When an unwanted spike appears in the dynamics of the slave before it does in the master, a control mechanism is triggered and sent to the master in order to prevent that event. The whole scheme is shown in Fig.~\ref{fig:scheme}. We now give some specific details of the method.

It is easy to see that the master system $x(t)$ is excitable for $|\mu|<1$ with fixed points at $x_{\pm}=\pm\arccos(\mu)$, and oscillatory for $|\mu|>1$. If the system is at the stable point $x(t)=x_{-}$ and a perturbation acts upon it for a short time, the way in which variable $x(t)$ returns to the stable value depends on the magnitude of the perturbation. Let us assume that we apply a single, delta-like, perturbation of magnitude $I_0$ at time $t_0>0$, i.e. we take $I(t)=I_0\delta
(t-t_0)$, when the system is at the stable fixed point $x(t_0)=x_{-}$. The effect of such delta function perturbation is to instantly change at $t_0$ the value of $x(t)$ such that $x(t_0^+)=x(t_0^-)+I_0$. The subsequent dynamics depends on the magnitude of $I_0$. If $I_0<2\arccos(\mu)$ the system returns smoothly to the fixed point $x_{-}$, whereas if $I_0 > 2\arccos(\mu)$ the variable $x(t)$ executes a full rotation (a spike) and returns to the equivalent point
$2\pi+x_{-}$. It is the aim of our control method to prevent those spikes. We do this by adding to the dynamics of the master, Eq.(\ref{ms1}), a small corrective signal $-\epsilon\delta(t-T_0)$ of magnitude $\epsilon$ at time $T_0$. The time $T_0$ is determined by two factors: a correction criterion and a response time. The correction criterion determines whether the corrective signal is sent, and the response time $t_R$ is the time it takes the correction action to act after it has been required.

\begin{figure}
\centering
\includegraphics[scale=0.25,angle=-90]{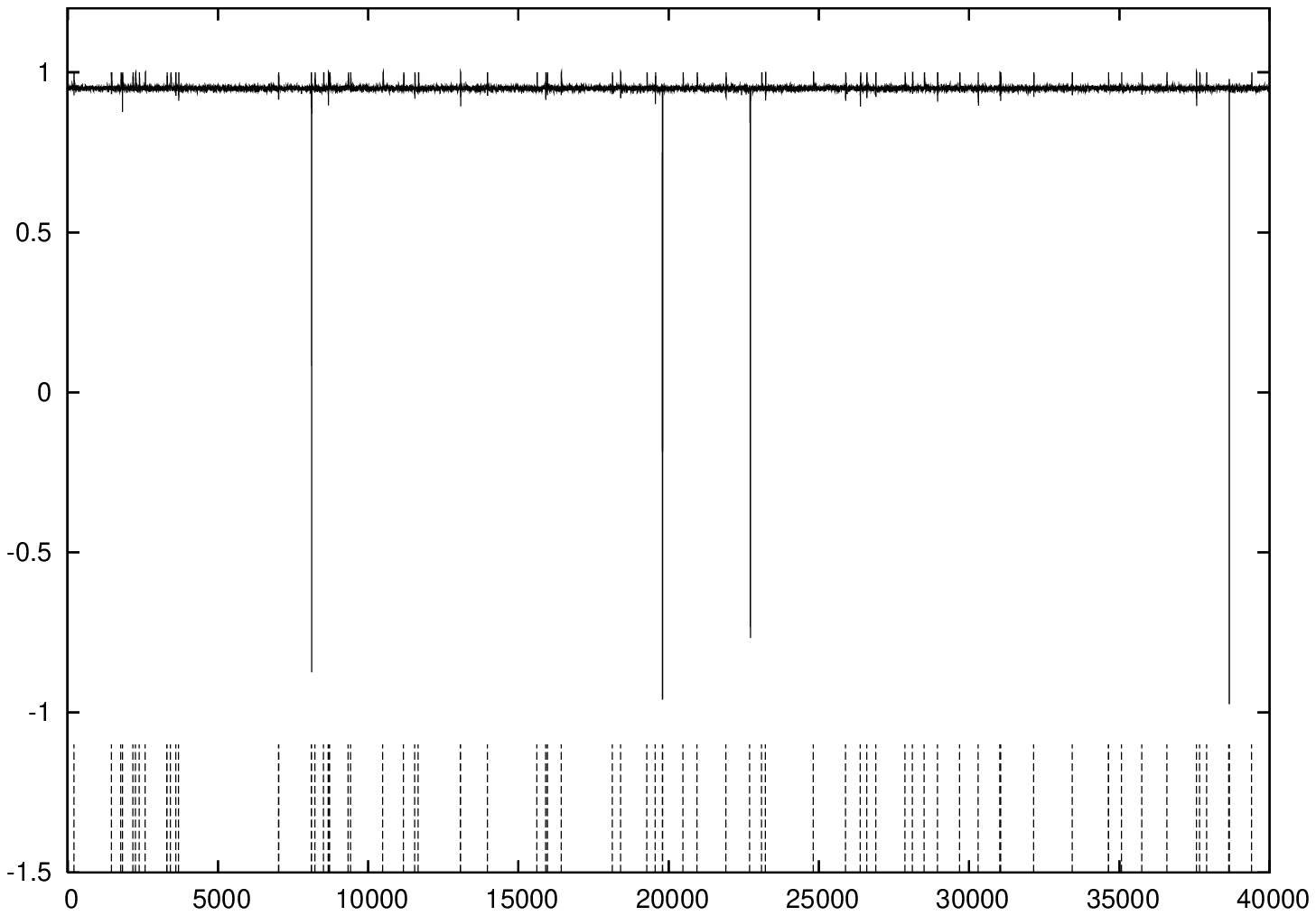}
\includegraphics[scale=0.25,angle=-90]{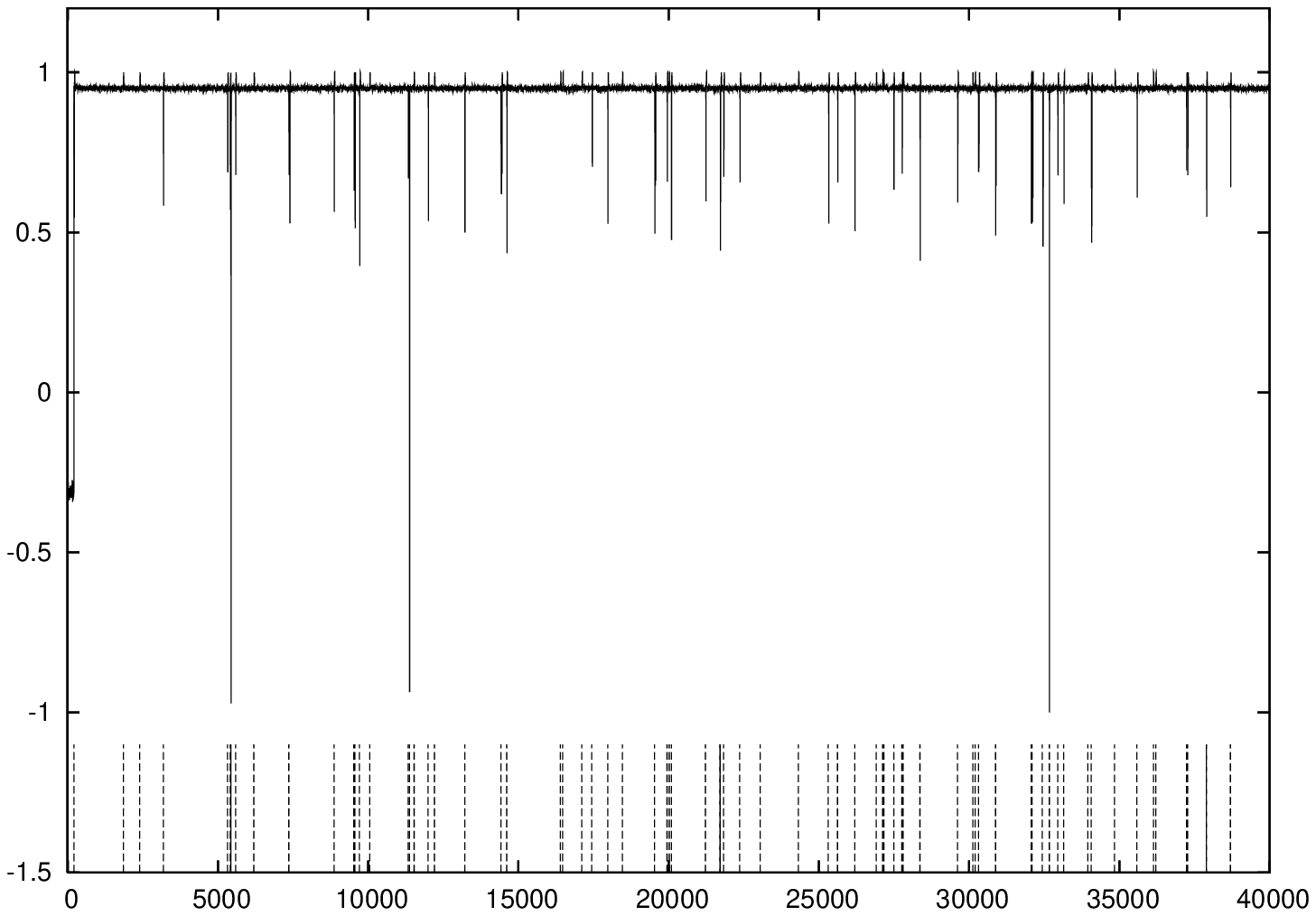}
\caption{\label{fig:many}Correction of a series of spikes. The external perturbation $I(t)$ consists of a series of random pulses (vertical lines al the bottom of both panels) following a Poisson distribution with $\langle t\rangle=500$ (there are $69$ excitations altogether). Noise of intensity $D=0.01$ has been added both to the master and to the slave. Top panel: control of magnitude $\epsilon=0.3$ using the slave variable $y$. Bottom panel: control of magnitude $\epsilon=0.85$ using the master variable $x$. The recovery time is $t_{rec}=5$ and the response time $t_R=1$, other parameters as in figure \ref{fig:single}.}
\end{figure}

Let us first consider a simple control method in which the correction criterion is that the variable $x(t)$ crosses a threshold value $x_0$. This resembles the DFC method except that we do not have to fix the delay time to stabilize a given orbit but the delay will correspond to the time it takes to inject a corrective signal into the master system when it crosses $x_0$, i.e. whenever $x(t)>x_0$. If this happens at time $t_1$, then at time $T_0=t_1+t_R$ the correction acts such that $x(T_0^+)=x(T_0^-)-\epsilon$, and it will be considered effective if $x(t)$ is brought back to the stable region. The sooner the correction action is taken, the smaller the magnitude of the correction $\epsilon$ needed. In any event, even for a zero response time, $t_R=0$, the condition $\epsilon > x_0-x_+$ must be satisifed in order to bring back $x(t)$ from the threshold value $x_0$ to the stable region. 

Our method works similarly but the decision of activating the corrective signal depends on the dynamics of the auxiliary system $y(t)$, coupled to $x(t)$ using the anticipated synchronization scheme discussed previously. The main idea is that the dynamics of $y(t)$ anticipates that of $x(t)$ and hence the correction criterion $y(t)>x_0$ happens earlier. Consequently, and this is an important advantage of the method, a much smaller magnitude $\epsilon$ is needed to suppress the undesired firing. Alternatively, a larger reaction time can be used to decide whether to apply the correcting signal. The correction criterion is now modified to the following: if $y(t)$ crosses the threshold value $x_0$ then a delta-like correction of magnitude $-\epsilon$ is applied to the variable $x(t)$. The same conditions for reaction and recovery times discussed before also apply here.
\begin{figure}
\centering \vspace{-1.8cm}
\includegraphics[scale=0.30,angle=0]{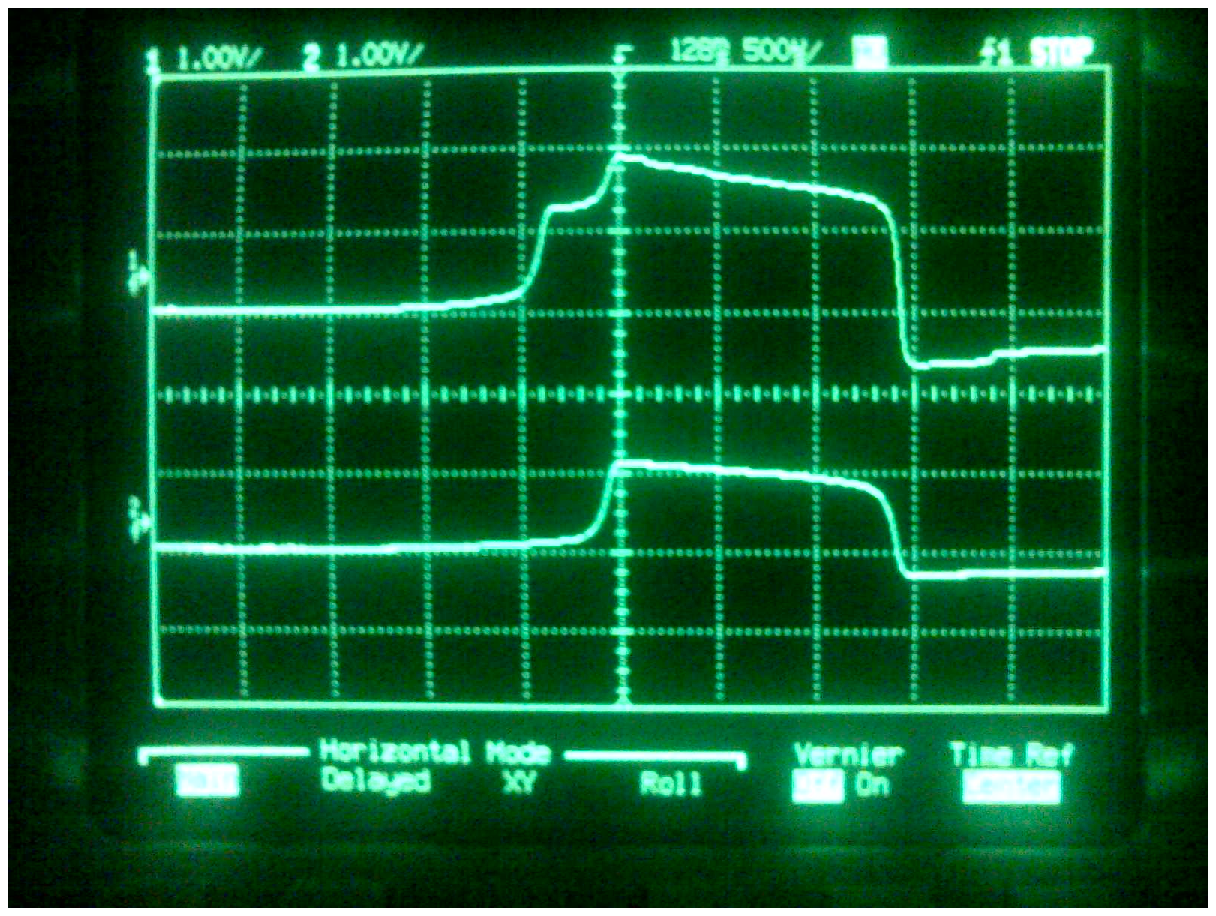}
\vspace{-1.8cm}
\caption{ \label{fig:exper1} Trajectories of the master (upper trace) and slave systems (lower trace) in a experimental implementation of the anticipated synchronization scheme in an electronic circuit simulating the FithHugh-Nagumo neuronal dynamics.}
\end{figure}

We now compare in detail the control methods using the criterions based upon the $y(t)$ or the $x(t)$ variables. When using, for example, a threshold value $x_0=\pi/4$,
our numerical results show that for a response time $t_R=0$, a value $\epsilon=0.468$ is needed when using the criterion on the $x(t)$ variable, while a much smaller value $\epsilon=0.022$ suffices in our scheme using the criterion based on $y(t)$. The same result applies in the case of a non-zero response time. If $t_R=1$, $\epsilon$ reduces from $0.834$, when using the criterion of $x(t)$, to $0.0307$ when using our method, or from $1.751$ to $0.0426$ if $t_R=2$. For $t_R=3$ the method based on $x(t)$ is unable to control the system while for our method it suffices to take $\epsilon=0.0597$. In Fig.~\ref{fig:single} we show that our control methods is efficient and indeed enables the suppression of a single spike.


In a typical control situation, the arrival of the perturbation and the consequent generation of spikes would be random. To model this, we consider a train of perturbations arriving at times $t_i$ such that the time intervals $t_{i+1}-t_i$ are distributed according to an exponential distribution of mean value $\langle tÊ\rangle$. We also add to both equations (\ref{ms1}) and (\ref{ms2}) independent white noise terms $D\xi(t)$ with correlations $\langle \xi(t)\xi(t')\rangle=\delta(t-t')$. We also consider that the control mechanism has a recovery time $t_{rec}$ such that $t_{rec}$ must elapse after a control signal has been applied before another control signal can be activated. The top panel of Fig.~\ref{fig:many} shows the result of our control method while the bottom panel shows the control method based on the $x(t)$ variable. It can be observed that not only the intensity $\epsilon$ of the control is smaller in the first case, but also the spikes are much better suppressed. Finally, we have checked that our control method still works well when the auxiliary system is not an identical copy of the master system.

\begin{figure}
\vspace{-1.8cm}
\includegraphics[scale=0.30,angle=0]{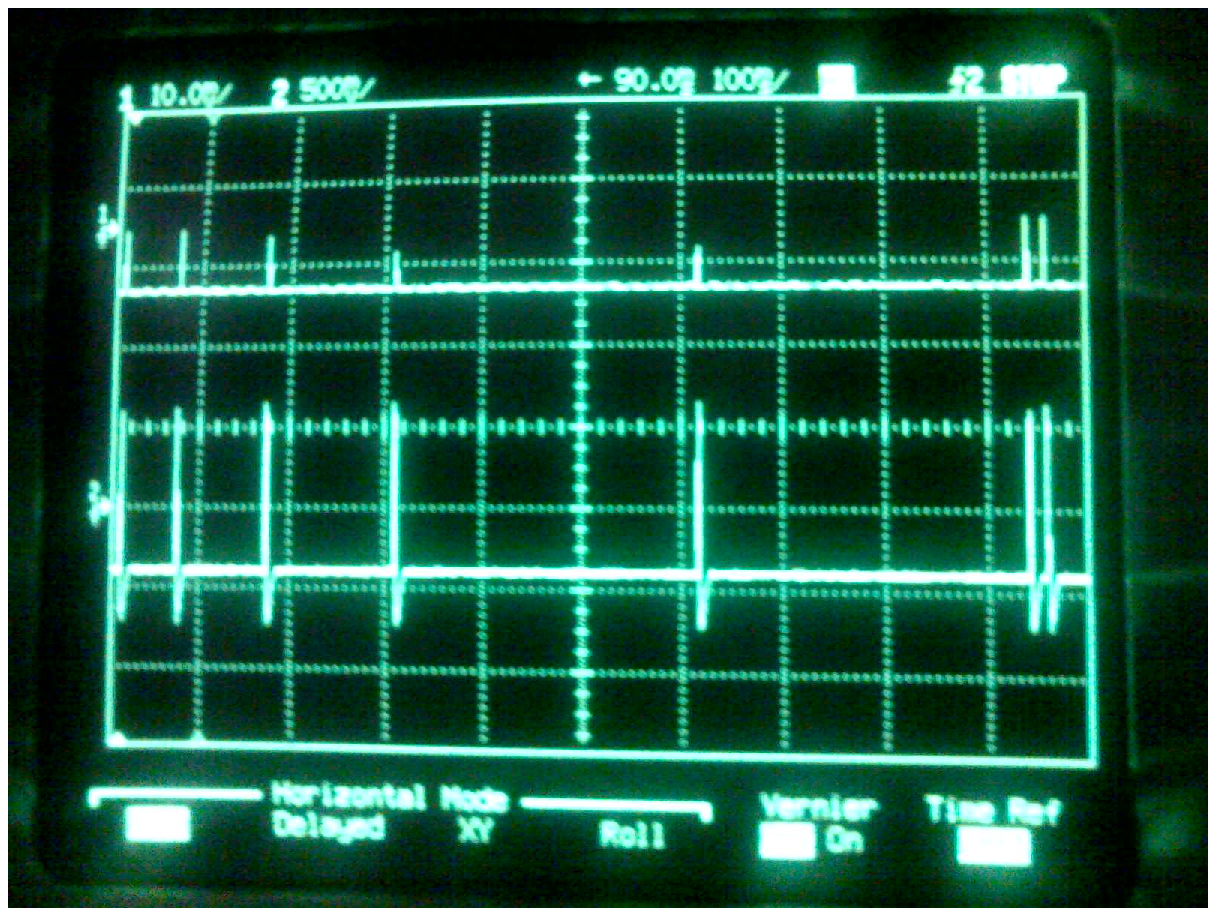}\\
\vspace{-3.6cm}
\includegraphics[scale=0.30,angle=0]{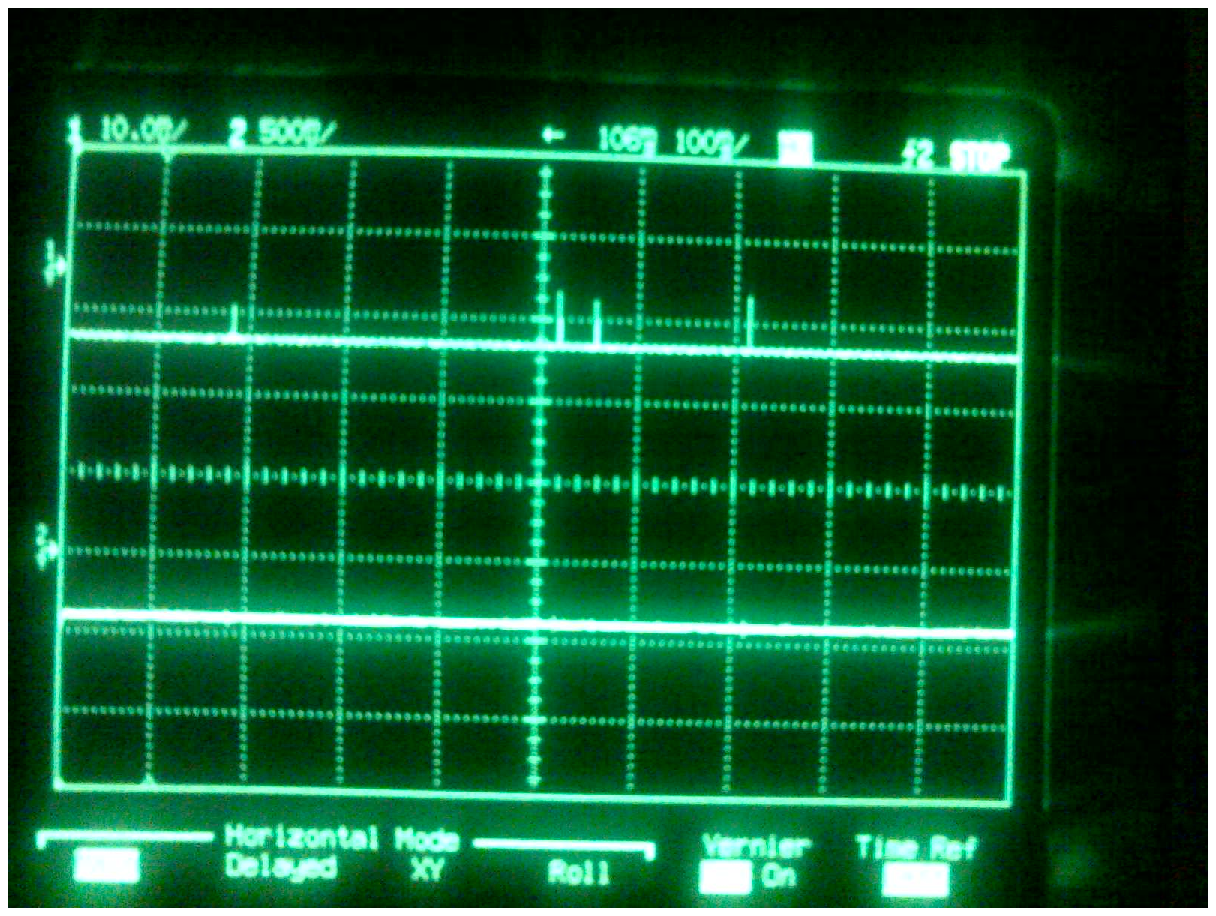}
\vspace{-2.cm}
\caption{ \label{fig:exper2} Experimental verification of the predict-control method in the electronic FithHugh-Nagumo system. The small vertical lines at the top of each figure are the external perturbations, and the lower traces are the master's variable. In the bottom panel, the spikes have been controled using a criterion based upon the dynamics of the slave (not shown). The top panel shows that the same procedure fails to prevent the spikes if the criterion is applied using the master's variable.}
\end{figure}

In order to assess the practical usage of our method we have implemented experimentally an electronic version of the FitzHugh-Nagumo neurons, another type of excitable system~\cite{fhn}. The details concerning the experimental setup can be found in~\cite{our}. For this work the set-up was modified introducing a monostable circuit that generates a pulse with programable variable width and delay when the slave system signal crosses a threshold. The pulse is applied as a corrective signal to prevent firings of the master. The complete diagram of the set-up can be found in~\cite{web}. Our experimental findings qualitatively agree with the numerical predictions. Fig.~\ref{fig:exper1} shows the dynamics of the master and the slave in response to a
perturbation applied only to the first. It can be clearly seen that the slave anticipates the firing of the master by roughly $400$ $\mu s$. In Fig.~\ref{fig:exper2} we show time traces taken from the oscilloscope. The top panel shows the results when the master system is controlled using its own output. In this situation it is almost impossible to cancel the firing. The bottom panel shows the results using our control method based on the anticipating slave. Note that all spikes have been suppressed in this particular realization of the experiment, a typical situation.

In conclusion, we have proposed a predict-prevent control method for the suppression of spikes in perturbed excitable systems. The method consists of two steps: (1) prediction made by the slave system with the use of the anticipated synchronization scheme and (2) correction applied to the master in order to suppress its unwanted response. We have demonstrated numerically and experimentally the efficiency of the method. The proposed technique has been applied to excitable systems but we hope that this work will stimulate additional studies in order to extend this method to other types of systems, specially chaotic ones, as well as to identify possible situations of practical interest.

We acknowledge financial support by the MEC (Spain) and FEDER (EU) through projects FIS2006-09966 and the EU NoE BioSim LSHB-CT-2004-005137.


\begin{thebibliography}{00}
\bibitem{control} Katsuhiko Ogata, {\sl Modern Control Engineering}, Prentice Hall (2001).
\bibitem{heart} A. V. Panfilov, Chaos \textbf{8}, 57 (1998).
\bibitem{ogy} E. Ott, C. Grebogi, J. A. Yorke, Phys. Rev. Lett. \textbf{64}, 1196 (1990).
\bibitem{pyragas} K. Pyragas, Phys. Lett. A \textbf{170}, 421 (1992).
\bibitem{sukow} J. Socolar, D. Sukow, D. Gauthier, Phys. Rev. E \textbf{50}, 3245 (1994).
\bibitem{luth} O. L\"{u}thje, S. Wolff, G. Pfister, Phys. Rev. Lett. \textbf{86}, 1745 (2001).
\bibitem{hall} K. Hall, D. J. Christini, M. Tremblay, J. J. Collins, L. Glas, J. Billete, Phys. Rev. Lett. \textbf{78}, 4518 (1997).
\bibitem{PRK01} A. Pikovsky, M. Rosenblum, J. Kurths, {\sl Synchronization: a universal concept in nonlinear sciences} Cambridge University Press (New York, 2001).
\bibitem{blekhman} I.I. Blekhman, {\sl Synchronization: in science and technology} (ASME Press, New York, 1988).
\bibitem{voss1} H. U. Voss, Phys. Rev. E {\bf 61}, 5115 (2000); Phys. Rev. E {\bf 64}, 039904 (2000); Phys. Rev. Lett. {\bf 87}, 014102 (2001).
\bibitem{chialvo} O. Calvo, D.R. Chialvo, V.M. Eguiluz, C. Mirasso, R. Toral, Chaos \textbf{14}, 1 (2004).
\bibitem{M01} C. Masoller, Phys. Rev. Lett. {\bf 86}, 2782 (2001).
\bibitem{HMM02} E. Hern\'andez-Garc{\'\i}a, C. Masoller,  C.R. Mirasso, Phys. Lett. {\bf A 295} 39 (2002).
\bibitem{our} M. Ciszak, O. Calvo, C. Masoller, C. Mirasso, R. Toral, Phys. Rev. Lett. \textbf{90}, 204102 (2003).
\bibitem{our2} R. Toral, C. Masoller, C. Mirasso, M. Ciszak, O. Calvo, Physica A {\bf 325}, 192 (2003).
\bibitem{KHTM05} M. Kostur, P. Hanggi, P. Talkner, J. L. Mateos, Phys. Rev. E \textbf{72}, 036210 (2005).
\bibitem{dynMar} M. Ciszak, F. Marino, R. Toral, S. Balle, Phys. Rev. Lett. \textbf{93}, 114102 (2004).
\bibitem{LTDASL02} Y. Liu, Y. Takiguchi, P. Davis, T. Aida, S. Saito, J.M. Liu, Applied Physics Letters, {\bf 80}, 4306 (2002).
\bibitem{voss2} H. U. Voss, Int. J. of Bif. and Chaos, {\bf 12}, 1619 (2002).
\bibitem{PJG08} A. N. Pisarchik, R. Jaimes-Reategui, J. H. Garcia-Lopez, Phil. Trans. Royal Soc. A {\bf 366}, 459 (2008).
\bibitem{mplb} M. Ciszak, R. Toral, C. Mirasso, Mod. Phys. Lett. B {\bf 18}, 1135 (2004).
\bibitem{adler} R. Adler, Proc. IRE {\bf 34}, 351 (1946); reprinted in Proc. IEEE {\bf 61} (1973). 
\bibitem{fhn} R. FitzHugh, Biophys. J. 1, 445 (1961); J. Nagumo, S. Arimoto, S. Yoshizawa, Proc. IRE {\bf 50}, 2061 (1962).
\bibitem{web} http://ifisc.uib-csic.es/claudio/activities.html
\end{thebibliography}
\end{document}